\documentclass[journal]{IEEEtran}

\usepackage{amsmath}
\usepackage{array}
\usepackage{epsfig}
\usepackage{graphicx}
\usepackage{amssymb}
\usepackage{color}
\usepackage{url}

\newcommand{\I}{\imath}
\newcommand{\diff}{\mathrm{d}}
\newcommand{\dg}{^{\mathrm{o}}}
\begin{document}

\title{Dictionary-free MR Fingerprinting reconstruction of balanced-GRE sequences.}

\author{Alessandro~Sbrizzi, Tom~Bruijnen, Oscar~van~der~Heide, Peter~Luijten and Cornelis~A.T.~van~den~Berg

\thanks{Part of this work was funded by the Dutch Technology Foundation (NWO-STW), grant number 14125.}
\thanks{Alessandro Sbrizzi, Tom Bruijnen, Oscar van der Heide, Peter Luijten and Cornelis A.T.  van den Berg are with the Center for Image Sciences, University Medical Center, Utrecht, Heidelberglaan 100, 3584 CX The Netherlands. e-mail:  a.sbrizzi@umcutrecht.nl.}
\thanks{Manuscript submitted to IEEE Transactions on Medical Imaging on the 4th of July 2017.}}

\maketitle 

\begin{abstract}Magnetic resonance fingerprinting (MRF) can successfully recover quantitative multi-parametric maps of human tissue in a very short acquisition time. Due to their pseudo-random nature, the large spatial undersampling artifacts can be filtered out by an exhaustive search over a pre-computed dictionary of signal fingerprints. This reconstruction approach is robust to large data-model discrepancies and is easy to implement. The curse of dimensionality and the  intrinsic rigidity of such a precomputed dictionary approach can however limit its practical applicability. 
In this work, a method is presented  to reconstruct  balanced gradient-echo (GRE) acquisitions with established iterative algorithms for nonlinear least-squares, thus bypassing the dictionary computation and the exhaustive search. The global convergence of the iterative approach is investigated by studying  the transient dynamic response of balanced GRE sequences and its effect on the minimization landscape. Experimental design criteria are derived which enforce sensitivity to the parameters of interest and successful convergence. The method is validated on simulated and experimentally acquired MRI data.

This manuscript was submitted to \textit{IEEE Transactions on Medical Imaging} on the 4th of July 2017.
\end{abstract}

\IEEEpeerreviewmaketitle

\section{Introduction}
\IEEEPARstart{T}{raditional} quantitative MRI techniques which aim at reconstructing the tissue relaxation times $T_1$ and $T_2$  rely on simple analytical signal models which can be fitted to a series of images taken at different time points \cite{qMRI}. In some special cases, the parameters of interest can be derived from analytical solutions of the model equation or by  solving a linear least-squares problem \cite{blueml,giri,despot}).\\
 When these straightforward approaches are not possible, nonlinear fitting algorithms can be applied to infer the parameters of interest \cite{schmitt,warntjes,looklocker}. A fundamental drawback of fitting nonlinear models is that the objective function could be non-convex. As a consequence,  the algorithm might converge to a local minimum, leading to erroneous parameter reconstructions. To circumvent this problem in the case when few unknowns are present in the model, brute force approaches can be applied; the measured signal is matched to a pre-computed dictionary of signal responses and the parameters combination with the highest correlation is retained \cite{fleyscher2007,fleyscher2008,bidhult}.\\
MR fingerprinting (MRF) \cite{ma2013} is a recent and innovative quantitative method which is based on a dictionary-match approach. In MRF, the exhaustive search over the pre-computed dictionary  makes it possible to derive  $T_1,T_2$ and proton density values ($\rho$) from images obtained with highly undersampled $k$-space. Initial work on MRF was based on gradient-balanced sequences, which are notoriously sensitive to off-resonance ($\omega$) \cite{hargreaves2001}. The discretization of $\omega$ has an impact on the accuracy of the fitting \cite{doneva2015,FISP_MRF} and, due to the increased computational and memory demands, can preclude the inclusion of other important parameters in the model such as the transmit field sensitivity \cite{cloos2016,buonincontri,ma2017}  and the slice profile effects \cite{ma2017}.  For these reasons,  gradient balanced acquisitions have to deal with fundamental practical issues and MRF-related research currently prefers spoiled sequences, which are much less dependent on off-resonance \cite{FISP_MRF}. Balanced sequences do have some important advantages, such as superior SNR and robustness to motion and flow \cite{hargreaves2012}.\\
 In this work, we consider balanced GRE sequence for MRF and cast the parameter reconstruction step as a nonlinear least-squares problem, which can be solved by established gradient-based iterative algorithms. We study the global convergence properties of such an iterative procedure and we derive rules to design balanced GRE sequences which lead to accurate  parameter estimations. Recent work on sequence design for MRF \cite{asslander2017} has focused on enhancing the sensitivity of the sequence to the parameters of interest within the dictionary-match framework. Our approach also addresses issues such as the presence of local minima and/or strongly irregular minimization landscapes which are crucial factors in dictionary-free iterative methods. The particular transient-state response of the obtained sequence makes it possible to  separate the true signal contribution from the undersampling artifacts. The corresponding objective function exhibits a  more regular profile than the ones obtained by standard MRF balanced GRE schemes and the convergence to the global minimum is thus facilitated.  The exhaustive search step can thus be avoided and the construction of the pre-computed dictionary, which could be affected by memory limitations and discretization issues especially along the $\omega$ direction, is no longer needed. Consequently, the signal model can be expanded and alongside $T_1, T_2, \rho$ and $\omega$, we also include the spatially varying radio-frequency (RF) field $B_1^+$, transceive phase $\phi$ and the slice profile variation throughout the transient-states sequence. \\
 We first illustrate the theory and rationale which pave the way to the design of so-called low-pass sequences. Afterwards, we show that the  dictionary-free  MRF framework is feasible by providing  numerical simulations and experimental results from a clinical 1.5 T scanner.
\section{Theory}
\subsection{MR fingerprinting reconstruction as nonlinear least-squares fitting}
 For balanced GRE sequences, the signal, $s$, emitted by a given homogeneous voxel, at position $\mathbf{r}$ at the $i$-th read-out is given by:
\[
 s_i(\mathbf{r}) = \alpha(\mathbf{r}) m(\beta(\mathbf{r}),t_i)
\]
where the variable $\alpha$ is the product of the  proton density ($\rho$) with the receive radiofrequency field amplitude ($B_1^-$) and with a phase term given by the transceive phase of the RF system ($\phi$). In short: $\alpha = \rho B_1^-\exp(\imath \phi)$. The remaining spatially dependent parameters are included in the variable $\beta$, with $\beta \equiv(T_1,T_2,  B_1^+, \omega)$. As shown in \cite{ma2017}, the transmit RF field amplitude $B_1^+$ must be included in the model to obtain accurate estimations. Finally, $m\equiv m_x+\I m_y$ denotes the transverse component of the nuclear magnetization in the rotating frame and is thus modelled by the Bloch equation \cite{jaynes}:
\begin{equation} 
\frac{\diff}{\diff t}\mathbf{m}=
\left(
 \begin{array}{ccc}
-\frac{1}{T_2} &\omega & -\gamma b_y(t) \\
 -\omega & -\frac{1}{T_2} &\gamma b_x(t)\\
 \gamma b_y(t) & -\gamma b_x(t) & -\frac{1}{T_1}
\end{array} 
\right)
\mathbf{m}+
\left(
\begin{array}{c}
0\\
0\\
\frac{1}{T_1}
\end{array}
\right),
\label{bloch}                     
\end{equation}
\[
\mathbf{m}(0)=
(0,0,1)^T\]
where $\mathbf{m} = (m_x,m_y,m_z)^T$ and $(b_x,b_y)$ are the transverse components of the applied radiofrequency field.\\
In this work, we are interested in the simultaneous estimation of $\alpha$ and $\beta$ from a dataset $d(t)$ by casting the reconstruction as a nonlinear least squares problem: 
\begin{equation}
\begin{array}{ccl}
 (\alpha^{\text{recon}}, \beta^{\text{recon}}) & = & \arg\min_{\alpha,\beta}\sum_{i=1}^{N}|\alpha m(\beta,t_i)-d(t_i)|^2 \\
 & = &\arg\min_{\alpha,\beta}\|\mathbf{M}(\beta)\alpha-\mathbf{d}\|^2\\
 & = &\arg\min_{\alpha,\beta}f(\alpha,\beta)
\end{array}
\label{ls}
\end{equation}
where $N$ is the number of acquisitions, $\mathbf{M}$ is the response of the Bloch equation in vector form and $\mathbf{d}$ the data vector. Equation (\ref{ls}) must be solved for each voxel to recover the spatially dependent parameters $\alpha$ and $\beta$.\\
The problem in Eq. (\ref{ls}) can be solved by a variable projection approach \cite{varpro}. Observing that, if $\beta^*$ is a solution of Eq. (\ref{ls}), then 
\begin{equation}
\alpha^* = \arg\min_{\alpha}f(\alpha,\beta^*) = \frac{\mathbf{M}(\beta^*)^H\mathbf{d}}{\|\mathbf{M}(\beta^*)\|^2}.
\label{lin}                                                                                                                                                              
\end{equation}
By substituting back into Eq. (\ref{ls}) we obtain the reduced functional:
\begin{equation}
 \beta^{\text{recon}} = \arg\min_{\beta}\left(\mathbf{I}-\frac{\mathbf{M}(\beta)\mathbf{M}(\beta)^H}{\|\mathbf{M}(\beta)\|^2}\right)\mathbf{d} 
\label{vp}
\end{equation}
which can be solved for $\beta$ by standard gradient-based iterative methods such as a trust-region algorithm. Note that the linearly dependent variable $\alpha$ no longer plays a role in the minimization problem and it can be recovered by solving  Eq. (\ref{lin}) after the solution of Eq. (\ref{vp}). The variable projection formulation of Eq. (\ref{vp}) has in general a faster convergence than the full problem, it is less susceptible to local minima if the function is not convex and it does not require initial guesses on $\alpha$. \\
The update step for the solution of Eq. (\ref{vp}) requires the Jacobian matrix of $f$ with respect to the nonlinear variables, that is, the components of $\beta$. To  calculate the derivatives, we employ forward mode automatic differentiation schemes \cite{ad} applied to the Bloch equation solver. In this way, the exact Jacobian can be obtained.
\subsection{Topology of the optimization landscape}
\begin{figure}[t!]
\centering
\epsfig{file=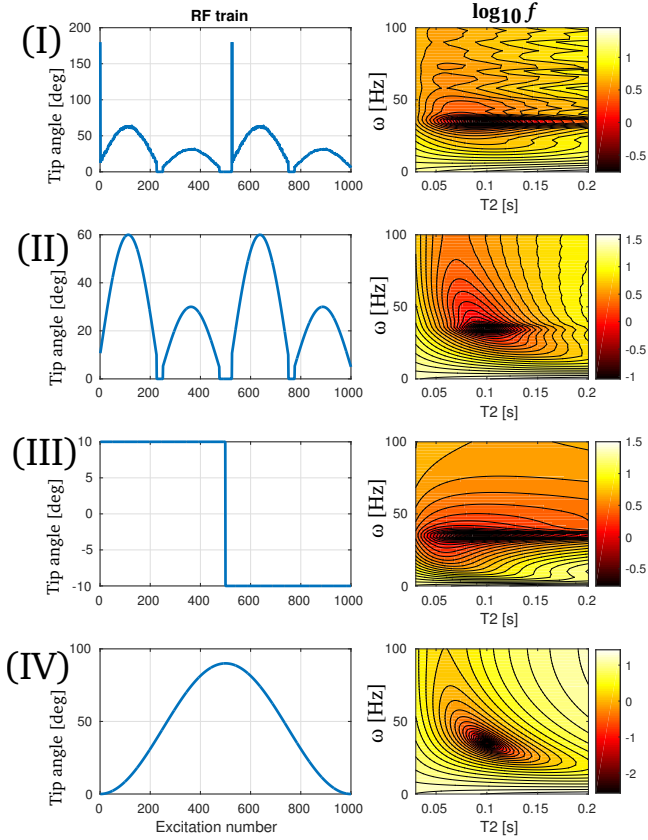, width=8.5cm}
\caption{The optimization landscapes in the $(T_2,\omega)$ plane for four different balanced GRE sequences. Note the several local minima for the randomized MRF sequence at the top (I) and the regularity of the landscape for the smooth sequence at the bottom (IV). }
\label{4landscapes}
\end{figure}

\begin{figure}[t!]
\centering
\epsfig{file=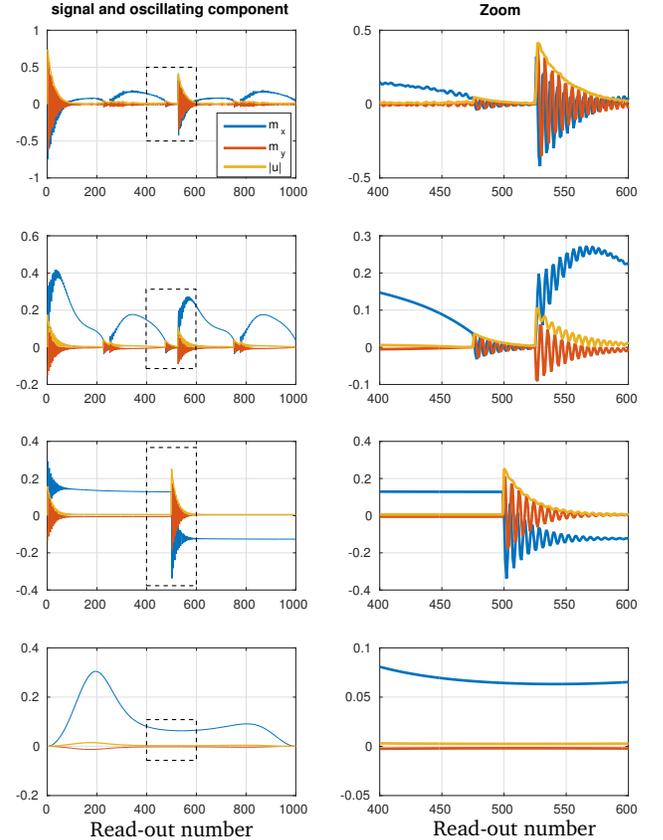, width=8.25cm}
\caption{The signal ($m_x,m_y$) and oscillatory components $|\mathbf{u}|$ for the four sequences from Fig. \ref{4landscapes}. The central section of each sequence is magnified in the right column of the figure. Note that $|\mathbf{u}|$ correctly quantifies the transient oscillatory behavior of the signal. }
\label{oscillations}
\end{figure}

Before we apply gradient-based methods for Eq. (\ref{ls}), we investigate its global convergence behavior by considering the topology of the optimization landscape. Since the reconstruction of $\alpha$ is a linear problem, thus convex, we focus on the topology of $f$ in the space of nonlinear dependent parameters, $\beta$. In particular, previous studies \cite{sbrizzi2017} reveal that for balanced sequences, the most irregular and possibly non-convex behavior is caused by the off-resonance response ($\omega$). For simplicity of exposition, we consider the simultaneous dependence of $f$ on $\omega$ and $T_2$. Similar behavior can be observed when $T_1$ and $B_1^+$ are also taken into account. \\
Figure \ref{4landscapes} shows four different balanced GRE sequences (left column) and the corresponding objective function $f$ (right column) plotted on a plane parallel to $(T_2,\omega)$ and passing through the global minimum. The global minimum corresponds to $(T_1,T_2,B_1,\omega)=(1.0\text{ [s]},0.1\text{ [s]},1.0\text{ [a.u.]},35\text{ [Hz]})$.  For all sequences, the number of excitations is $N = 1000$ and the echo and repetition times are constant and set to   $(T_E,T_R)=(2.4,4.8)$ ms.  \\
Sequence I is similar to the MRF acquisition used in \cite{ma2013} and represents a typical randomized MRF acquisition: note the small random component superimposed to the sinusoidal lobes and the two inversion pulses. Sequence II is similar to the first one but neither random perturbations nor inversion pulses are employed. Sequence III is characterized by a piecewise constant tip angle and the RF train of sequence IV is given by $90\dg \sin^2 (j\pi/N)$ with $j=1,\dots,N$. \\
The real and imaginary signal components, that is, $m_x$ and $m_y$ are shown in Fig. \ref{oscillations}. From Figures \ref{4landscapes} and \ref{oscillations} we observe that:
\begin{itemize}
 \item the randomized MRF sequence (I) exhibits a strongly non-convex profile, including several local minima; 
\item the signals from the sequences I, II and III exhibit high frequency oscillatory components, which appear in concomitance with discontinuities (jumps) in the tip angle train;
\item the regularity of the minimization landscape is strongly related to the smoothness of the signal. The smoother the signal, the more regular the corresponding landscape; in particular, sequence IV gives rise to a very regular,  basin-shaped landscape.
\end{itemize}

The application of derivative-based iterative algorithms for solving  Eq. (\ref{ls}) may lead to suboptimal solutions  and thus wrong parameter values when sequences  I-III are employed. This is one of the main reasons why MRF reconstructions are carried out by an exhaustive search over the whole (discretized) space of possible solutions (dictionary). Unfortunately, the dictionary-based approach suffers from discretization and memory limitations and it may even become infeasible and/or inaccurate when increasing the dimensionality of the model. Furthermore, adaptation of sequence parameters requires re-computation of the dictionary; impairing the flexibility of the exhaustive search approach. Since the  smooth response (sequence IV) gives rise to a well behaved objective function, we expect these kind of excitation schemes to lead to fast and global convergence when iterative minimization methods are applied. \\
Reconstruction in the presence of noise requires special attention since the performance of iterative algorithms could worsen due to the mismatch between model and data. MRF signals typically exhibit large high-frequency components which originate from two sources: incoherent $k$-space undersampling  and inherently high-frequency response of the MRF sequence itself. The latter is illustrated at the top  of Fig. \ref{oscillations} where the  signal  for the MRF sequence from \cite{ma2013} is plotted \textit{without} undersampling artifacts. Intuitively, the inherently high frequency components in the response of sequences I-III make it difficult to separate the aliasing artifacts from the true signal evolution, since both model as well as residual  are expected to incorporate a large noise-like term. On the other hand, if the true signal evolution (Bloch equation response) were confined in a purely low-frequency band, a correct signal/artifact separation should be attainable;  after all, the irregular $k$-space undersampling scheme would still guarantee that the corresponding artifacts consist of mainly high frequency behavior. 
In this  case, the Bloch-based least-squares fitting (Eq. \ref{ls}) would automatically act as a low-pass filter and we name these types of sequences ``low-pass balanced-GRE sequences''. Note the analogy of this approach with the compressed sensing technique \cite{cs}, whose fundamental requirement is the incoherence between the sampling  and the sparsity representations; the incoherence allows for the correct filtering of artifacts hence accurate recovering of data also in the case of strongly undersampled scenarios.
 \subsection{Low-pass balanced GRE sequences}
In this paragraph, we investigate the transient response of balanced GRE sequences and provide conditions to design RF trains which are characterized by an inherently smooth signal behavior. Our analysis is inspired by the work of Hargreaves \cite{hargreaves2001}.\\
Consider the Bloch equation (\ref{bloch}) where, without loss of generality, the RF excitations are applied along the $y$-axis, that is, $b_x=0$ and $b(t) = b_y(t)$. 
The Bloch equation is nonlinear in the RF pulse and, for typical MRF sequences where the magnetization is kept in the transient states, non-stationary. In general, the study of nonlinear, non-stationary differential equations is a challenging subject of dynamical systems theory. To simplify the analysis, we neglect the effects of relaxation between two consecutive read-out moments $t_j$ and $t_{j+1}=t_j+T_R$. Furthermore, we apply an averaging step to the RF excitation train by representing it as a piecewise constant function such that $b(t_j) = b_j$ for $t\in[t_j,t_{j+1})$ and 
\[b_j=\frac{1}{T_R}\int_{t_j}^{t_{j+1}}b(t)\diff t,\quad j = 0,\dots,N-1.\] 
The effect of $b(t_j)$ and $\omega$ on $\mathbf{m}_j\equiv\mathbf{m}(t_j)$  can thus be  approximated by a rotation, $\mathcal{R}_j$:
\[
 \mathbf{m}_{j+1}\approx \mathcal{R}_j\mathbf{m}_j
\]
 where the rotation axis, $\mathbf{v}_j$, and the rotation angle, $\phi_j$, are given by  
\[
 \mathbf{v}_j =  \frac{1}{\sqrt{\gamma^2b_j^2+\omega^2}}
\left(\begin{array}{c}
 0\\
\gamma b_j\\
\omega
\end{array}\right),\quad\phi_j = T_R\sqrt{\gamma^2b_j^2+\omega^2}.
\]
If $\mathbf{m}_j$ is parallel to $\mathbf{v}_j$ there is no change in the magnetization, thus $\mathbf{m}_{j+1} \approx \mathbf{m}_j$. The oscillatory behavior is determined by the rotating component of $\mathbf{m}_j$, which lies on the plane orthogonal to $\mathbf{v}_j$ and is denoted by $\mathbf{u}_j$: 
\[\mathbf{u}_j\equiv (I-\mathbf{v}_j\mathbf{v}_j^T)\mathbf{m}_j.\]
Clearly, the projection of the rotating $\mathbf{u}_j$ onto the transfer plane will manifest itself as an oscillatory motion. The larger $|\mathbf{u}_j|$, the larger the oscillations will be.
 To illustrate this fact, Fig. \ref{oscillations} reports the signal (i.e. the transverse components of $\mathbf{m}_j$) and the amplitude of the oscillatory component, $|\mathbf{u}_j|$, for the four sequences already shown in Fig. \ref{4landscapes}. The right column is a magnification of the central part of each sequence. Note that $\mathbf{u}_j$ correctly estimates the magnitude of the transient oscillatory component of the signal. As already observed in the previous paragraph, each jump of the RF train, including the one at the beginning, causes a sudden increase in the oscillation's amplitude; the rotation axis undergoes an abrupt change in inclination, causing the magnetization to acquire a component in the oscillatory plane.\\
 For the smooth sequence starting with $b_0\approx 0$ (IV) the response remains smooth throughout all read-outs.  At $t=0$,  $\mathbf{m}$ is in the static equilibrium, that is, $\mathbf{m}_0 = (0,0,1)^T$;  the oscillatory component is minimized if the rotation axis, $\mathbf{v}_0\propto(0,\gamma b_0,\omega)$ lies predominantly along the $z$ direction. This explains why  the RF train which starts with $b_0\approx 0$ does not cause oscillations at the beginning. By slowly increasing the tip angle, the rotation axis starts to precess around the $x$-axis and the magnetization follows at a small angle with $\mathbf{v}$. The smooth behavior of $b$ implies small variations in the rotation axis and, by consequence, it ensures that $\mathbf{m}$ remains locked on to $\mathbf{v}$; the oscillatory components are thus minimized throughout the whole sequence. Note the analogy with the dynamics of adiabatic half passage pulses \cite{bendall1986}, although they act on a much smaller time scale. \\
The  case of $\omega \approx 0$ Hz deserves special attention. In this situation, the rotation axis is always directed along $y$  and oscillations occur if $\mathbf{m}$ has a nonzero longitudinal component, which is certainly the case at the beginning of the sequence.  Balanced sequences are usually implemented however with a $180\dg$ RF phase cycling to avoid signal cancellation  and the signal response in this case is shifted by $1/2T_R$ Hz  in the frequency band \cite{hargreaves2001}. The oscillatory condition becomes $\omega \approx \pm   1/2T_R$ Hz and, given the short repetition times of MRF acquisitions, we can assume that these off-resonance frequency bands are not present. 
\subsection{Experimental design}
Having shown that a smooth RF train is characterized by well-behaved objective function, we proceed by designing a low-pass sequence whose response is also highly sensitive to the parameters of interest, namely $\rho$, $T_1$ and $T_2$.\\
 Inference based on (nonlinear) least-squares leads to the quantification of a problem's sensitivity with respect to the parameters. In particular, the covariance matrix of the problem carries important information since its diagonal entries can be used to estimate the standard deviation of the reconstructions. Denoting the covariance matrix by $\mathbf{C}$, we have that 
$\mathbf{C}\approx\eta^2( \mathbf{J}^T\mathbf{J})^{-1}$ where $\mathbf{J}$ is the Jacobian matrix of the model and $\eta$ is the  noise variance. The $(i,j)$-th component of $\mathbf{J}$ is given by 
\begin{equation}
 [\mathbf{J}]_{i,j} = \frac{\partial \alpha m(\beta,t_i)}{\partial \xi_j}
\label{cov}
\end{equation}
where $\xi_j$ is one of the six parameters of the model, namely  $\alpha^R,\alpha^I, T_1,T_2,B_1^+,\omega$, The complex  variable $\alpha$ is split  into its real and imaginary parts, that is $\alpha^R=\Re(\alpha)$ and $\alpha^I=\Im(\alpha)$.
The standard deviation of the $n$-th parameter is finally given by $\sigma_n\approx\sqrt{[\mathbf{C}]_{n,n}}$ and it should be minimized for each parameter of interest. This is a standard approach in experimental design and it has recently been applied to MRI sequence optimization  \cite{asslander2017,teixeira}.\\
To obtain a balanced-GRE sequence which is simultaneously low-pass and sensitive to the parameters of interest, we solve the optimal experimental design problem:
\begin{equation}
\begin{array}{rl}
 \min_{\boldsymbol{\theta}} & \max_n\left\{\frac{\sigma(T_1^n)}{T_1^n},\frac{\sigma(T_2^n)}{T_2^n},\frac{\sigma(\rho^n)}{\rho^n}\right\} \\
 \text{such that:} & \theta_0 \leq \epsilon \\
                   & \|D\boldsymbol{\theta}\|_{\infty}\leq\delta\\
                   & \|\boldsymbol{\theta}\|_{\infty}\leq \theta_{\max}
\end{array}
\label{design}
\end{equation}
where $\boldsymbol{\theta}$ denotes the  tip angles train, $D$ is a first order finite difference operator and the standard deviation $\sigma$ is calculated for a test set of target parameters. For this purpose, we choose 25 different values of $(T_1,T_2,B_1^+,\omega)$ which are randomly drawn in the range:
\[
 \begin{array}{rcl}
  T_1 &\in & [250,3000 ] \text{ ms}\\
T_2 &\in & [50,250 ] \text{ ms}\\
B_1^+ &\in & [0.9,1.1] \text{ a.u.}\\
\omega &\in & [-150,150]\text{ Hz}.
 \end{array}
\]
As shown in the previous section, high-frequency oscillatory response is averted by assigning a very small value to the first tip angle $\theta_0$ (first constraint in Eq. (\ref{design})) and by forcing the tip angles to slowly change  (second constraint). For instance, we set $\epsilon = 1\dg$ and $\delta$ equivalent to twice the value achieved by  sequence IV of Fig. \ref{4landscapes}.  The third constraint limits the RF peak power and we set $\theta_{\max}=180\dg$. No inversion pre-pulses are employed. In the optimization, $\boldsymbol{\theta}$ is parametrized in terms of 10 cubic spline functions, which ensure that the tip angle train is very smooth over the whole sequence. The design problem is thus solved for the 10 spline coefficients. \\
Other sequence parameters are: $(T_E,T_R) = (2.4,4.8)$ ms, 1000 RF excitations. The RF pulse duration is 1 ms. Optimization is carried out with a Matlab built-in interior-point algorithm, called through the function \texttt{fmincon} and stopped after 100 iterations. As a smooth initial guess, sequence IV from Fig. \ref{4landscapes} was selected. In all tests reported in this work, the forward signal simulations are carried out with the C language implementation of the Bloch simulator by Brian Hargreaves \cite{blochsim}.\\
\begin{figure}[t!]
\centering
\epsfig{file=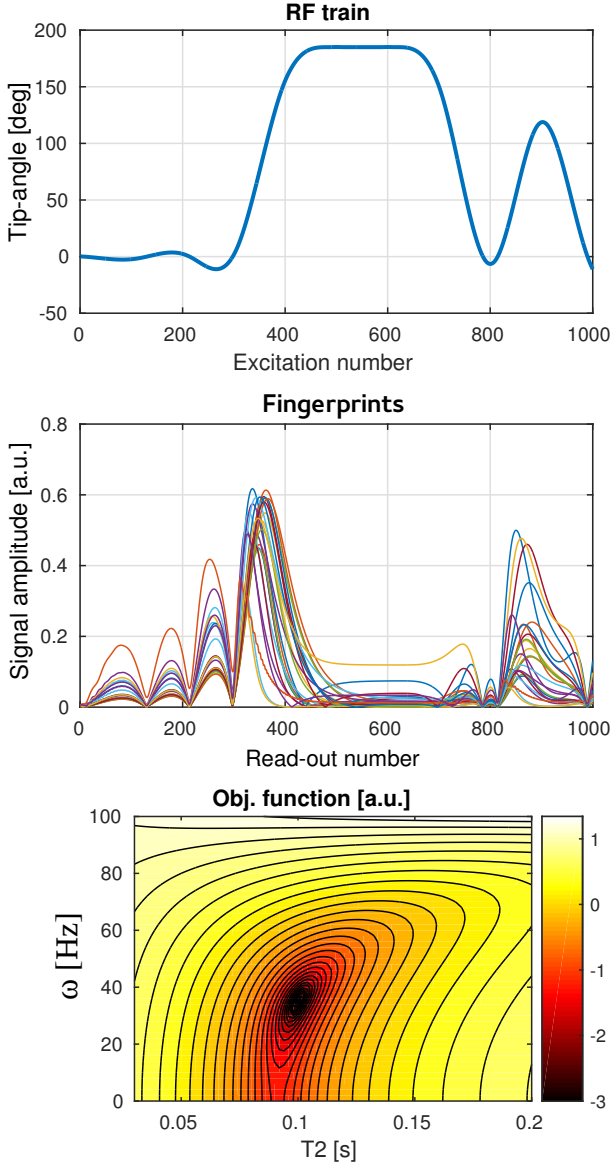, width=8cm}
\caption{The optimized low-pass sequence obtained upon solving problem (\ref{design}). Top: the tip angle train. Middle: the signal fingerprints for the 25 test parameter values. Note the smooth behavior throughout the whole sequence. Bottom: the optimization landscape for the case $(T_1,T_2,B_1^+,\omega)=(1.0\text{ [s]},0.1\text{ [s]},1.0\text{ [a.u.]},35\text{ [Hz]})$ reveals a regular, basin-like profile.}
\label{lowpass_seq}
\end{figure}
The low-pass sequence obtained upon the solution of problem (\ref{design})  is shown in Fig. \ref{lowpass_seq}.  The corresponding signal fingerprints for all 25 test parameter values are reported in the middle plot and confirm the smooth signal behavior throughout the whole sequence. The optimization landscape for $(T_1,T_2,B_1^+,\omega)=(1.0\text{ [s]},0.1\text{ [s]},1.0\text{ [a.u.]},35\text{ [Hz]})$ is shown at the bottom. As expected, the objective function is characterized by a regular, basin-like profile. 
\section{Multi-parametric reconstructions}

\subsection{Simulated acquisition}
\begin{figure}[t!]
\centering
\epsfig{file=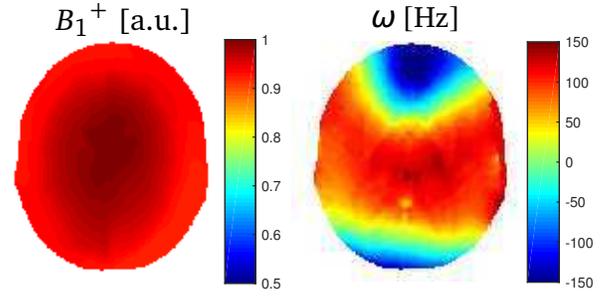, width=8cm}
\caption{The radiofrequency field ($B_1^+$) and off-resonance ($\omega$) maps for the simulated reconstruction test.}
\label{b1b0}
\end{figure}

\begin{figure*}[t!]
\centering
\epsfig{file=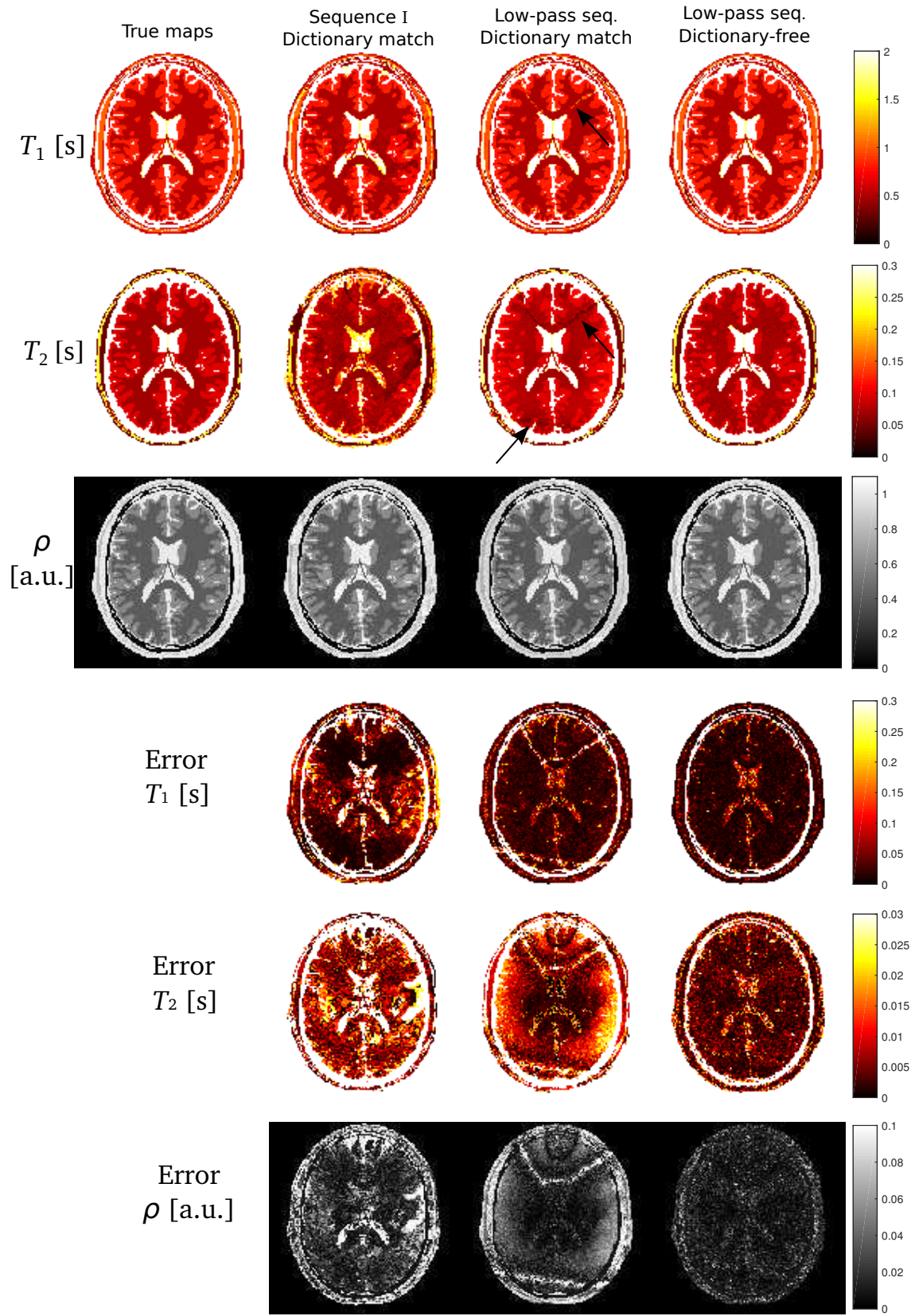, width=15cm}
\caption{$T_1$, $T_2$ and protons density images reconstructed from data corresponding to sequence I \cite{ma2013} and the optimized low-pass sequence (Fig. \ref{lowpass_seq}). The error maps are reported in the lower half of the figure. Reconstructions are performed with dictionary matching (second and third column) and with the nonlinear least squares approach (fourth column). The matching reconstructions suffer from dictionary discretization problems, indicated by the arrows in the $T_2$ map of columns two and three. The dictionary-free reconstruction (fourth column) resolves the discretization problem. }
\label{images}
\end{figure*}

Once the optimal sequence is obtained, we simulate the reconstruction of a $128\times128$ digital brain phantom \cite{brainweb} at 1.5 T, which is reported on the left of Fig. \ref{images}. The off-resonance and RF transmit field variations are scaled to be in the range of, respectively, $\omega\in[-150,+150]$ Hz and $B_1^+\in[0.9,1.0]$ a.u. and are shown in Fig. \ref{b1b0}.

The $k$-space sampling schemes employed for MRF acquisitions are usually based on spiral or golden-angle radial acquisitions to enhance pseudo-randomness of the undersampling artifacts in the temporal domain. In this test, the artifacts are simulated as 0-mean, normally distributed random perturbations $p_i$ ($i = 0,\dots,N-1$) and are scaled to obtain an SNR = 3 level, which is defined as
\[
 \text{SNR}=\frac{\|\text{signal}\|_2}{\|\text{perturbations}\|_2}=\frac{\sqrt{\sum\limits_{i=0}^{N-1}|s_i|^2}}{\sqrt{\sum\limits_{i=0}^{N-1}|p_i|^2}}=3.
\]
No slice profile variation is taken into account for the simulation.\\
All voxels from the corrupted brain images are reconstructed  following the proposed dictionary-free nonlinear least-squares approach, Eq. (\ref{ls}), by applying the variable projection method \cite{varpro} and a trust-region algorithm to the complex signal of the low-pass sequence. We estimate the proton density $\rho=|\alpha|$, $T_1$ and $T_2$ while the remnant unknowns ($B_1^+,\omega,\phi=\angle \alpha$) are treated as nuisance parameters, that is, they are reconstructed but their estimation is not required to be precise. As initial estimates we set $(T_1,T_2,B_1^+,\omega)^{\text{start}}=(0.75\text{ [s]}, 0.1\text{ [s]}, 1.0 \text{ [a.u.]}, 0 \text{ [Hz]})$. The algorithm is halted when the relative change in the objective function between two consecutive iterations is smaller than  $10^{-6}$. 
 As a comparison, we perform a dictionary match reconstruction for the same low-pass sequence data  and for the typical MRF sequence  \cite{ma2013} (sequence I in Fig. \ref{4landscapes}) with the equivalent number of excitations and $(T_E,T_R)$ values.  The dictionary entries for these two exhaustive search reconstructions are evaluated  in the intervals $T_1\in[0.263,2.867 ]$ s, $T_2\in[0.037,0.401]$ s with 5\% relative increase steps and $\omega\in[-150,150]$ with 5 Hz steps. Given the memory limitations,  $B_1^+$ is not included in the dictionary but is assumed to be everywhere equal to 1.\\
The parameter maps reconstructed with the dictionary match approach and the proposed nonlinear least-squares method are shown in, respectively, the third and fourth column of Fig. \ref{images}. The second column of the same figure shows the dictionary match reconstruction from sequence I  \cite{ma2013}. Note that the reconstructed values from the low-pass sequence are more accurate. Furthermore, the maps reconstructed with the proposed dictionary-free method show higher accuracy than the exhaustive-search reconstructions. This can be appreciated by looking at the error maps (Fig. \ref{images}-bottom). Clearly, the inaccuracy of the dictionary-based reconstructions is caused by the exclusion of the $B_1^+$ from the model and the relatively coarse discretization of $\omega$. 
\subsection{Experimental validation with a 1.5 Tesla MRI scanner}
\begin{figure}[t!]
\centering
\epsfig{file=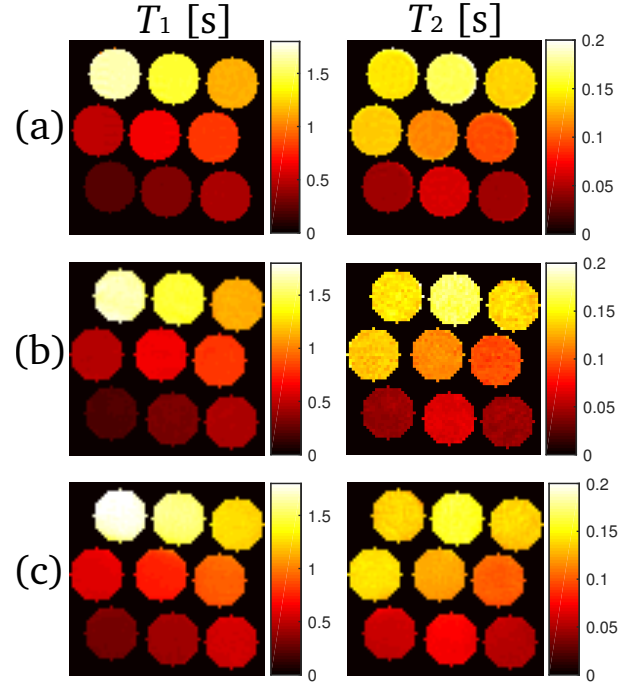, width=8cm}
\caption{Experimental validation of a low-pass sequence with nine gel phantoms with varying relaxometric properties. (a) $T_1$ and $T_2$ maps as obtained from the reference inversion-recovery spin-echo protocol; (b) parameter maps obtained from the optimized low-pass sequence with 21 spokes per time-point using the dictionary-free approach; (c) same as in (b) with 3 spokes. }
\label{phantoms}
\end{figure}

\begin{figure}[h!]
\centering
\epsfig{file=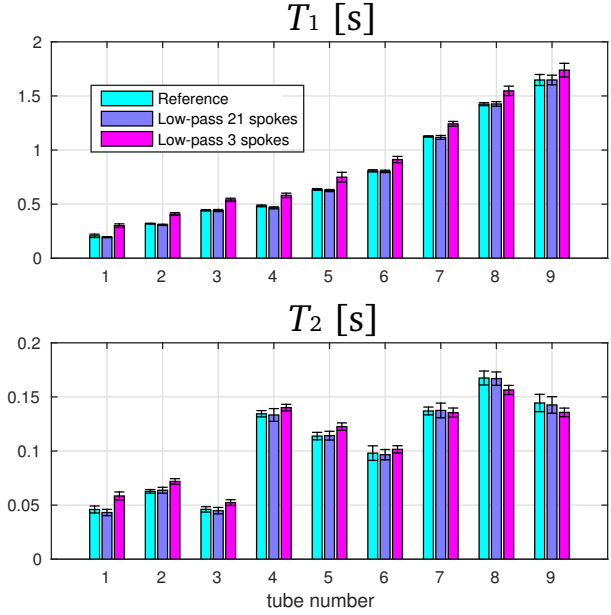, width=8cm}
\caption{Histogram plots  for the experimental validation test. Note that even at extremely high undersampling rate of 66 (i.e. 3 spokes), the dictionary-free low-pass reconstruction is rather accurate.}
\label{bars}
\end{figure}

\begin{figure}[t!]
\centering
\epsfig{file=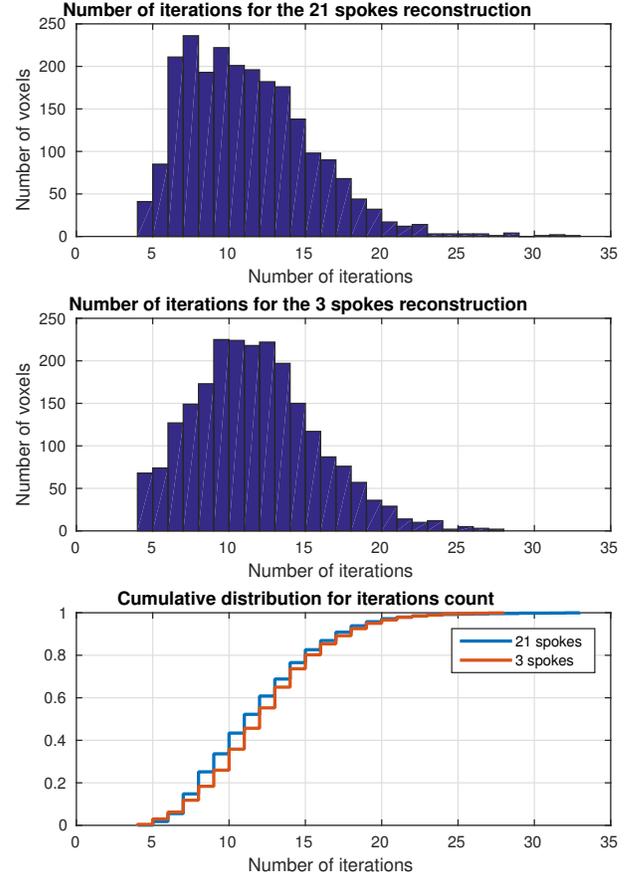, width=8cm}
\caption{The number of iterations needed for the iterative nonlinear least square reconstruction applied to the experimentally recorded data. Top: histogram distribution for the 21 spokes acquisition. Center: histogram distribution for the 3 spokes acquisition. Bottom: empirical cumulative distribution function for both acquisitions. Relatively few iterations are needed by the trust-region algorithm to converge. }
\label{iterations}
\end{figure}

Experimental validations of the low-pass dictionary-free reconstruction were performed with a 1.5T scanner (Philips-Ingenia, Best, The Netherlands) using a 16 receive channels head-coil and nine cylindrical gel phantoms  with different relaxometric $(T_1,T_2)$ properties (TO5, Eurospin II test system, Scotland). A 2D golden-angle radial $k$-space trajectory \cite{winkelmann} was employed where the  RF train was repeated for the multiple spokes per time-point as in \cite{cloos2016}. Each RF train was separated by a  5 second pause to allow for total relaxation. The aliased magnetization images at each time-point (snapshots) were reconstructed using iterative SENSE \cite{pruessmann2001} with coil sensitivity maps estimated as in \cite{uecker2014} and the non-uniform fast Fourier transform implementation from \cite{nufft}. The dictionary-free reconstruction  was subsequently performed on a voxel-by-voxel basis as described in the previous section. To take the slice profile variation throughout the RF train into account \cite{ma2017}, the response of the RF pulse was evaluated and integrated also in the slice-selection direction. A Gaussian-shaped RF pulse of 2.0 ms duration was used in combination with a slice selective gradient to obtain a 5 mm slice thickness, which is defined as the full-width-half-maximum (FWHM) value. The RF pulse was defined on a temporal grid with 0.1 ms steps while the slice profile direction was discretized over a 20.0 mm range and 1.0 mm steps.\\
 The in-plane resolution was $1.2\times1.2$ mm$^2$ and the size of the reconstructed matrix was $128\times 128$. Two undersampled reconstructions were considered, one with 21 spokes (effective reduction factor $\approx 9.5$) and one with 3 spokes (effective reduction factor $\approx 66$). \\
For validation purposes, we also acquired (a) a standard inversion-recovery spin-echo protocol (10 inversion times with values in $[0.1,3.0]$ ms) and (b) a single echo spin-echo protocol (10 echo times with values in $[0.2, 5.0]$ s) with $T_R=10$ s for $T_1$ and $T_2$ mapping, respectively.\\
The $T_1$ and $T_2$ maps reconstructed from the scanner measurements are shown in Fig. \ref{phantoms}. The mean and standard deviation values calculated over each tube are reported in Fig. \ref{bars}. The low-pass dictionary-free estimation of the relaxation rates agrees well with the values from the spin-echo protocol. Even with the  high undersampling factor of 66 (3 spokes), the quantitative value of $T_1$ and $T_2$ are close to the reference scan. \\
Finally, the number of iterations needed by the trust-region algorithm to reconstruct the experimentally acquired data are reported in Fig. \ref{iterations}. For both acquisitions, the number of iterations is rather small; on average, the 21-spokes and the 3-spokes reconstructions required, respectively, 11.8 and 12.3 iterations. Overall, no more than 35 iterations were needed and 95\% of the voxels were reconstructed with less than 20 iterations (see the cumulative distribution diagram at the bottom of Fig. 
\ref{iterations}).  The regular-shape of the objective function for the low-pass sequence leads to fast convergence of the dictionary-free approach.

\section{Discussion}  
In this work, we have shown that the reconstruction of MR fingerprinting data can be implemented as a solution of a nonlinear least squares problem using established iterative, gradient-based algorithms. The advantage of this approach is that the computation of a large-scale dictionary is no longer needed, enabling straightforward extensions of the signal model to incorporate multiple parameters. Although novel strategies are being explored for the acceleration of the dictionary computation and matching \cite{yang2017}, such a brute-force approach can only be successful and practical for a small number of parameters in the model. The model used in this work relies on six unknowns but this could be extended to include effects such as partial volumes, magnetization transfer and diffusion. Additionally, since each change in the sequence requires a new dictionary construction, flexibility is enhanced when a dictionary free method is applied.\\
A dictionary-free approach has recently been presented in \cite{zhang2017} by application of a Kalman filter but only shows results for extremely simplified  simulations, where important variables such as the proton density, the transmit profile $B_1^+$ and the slice profile variation throughout the sequence are not taken into account. Furthermore, only a very small off-resonance range is considered. Another limiting factor for that approach is that the successful application of the Kalman filter is strongly dependent upon the algorithmic parameters such as the choice of the covariance matrix which models the process noise. In our own experience, this is not a trivial step for data which in general is not Gaussian distributed and whose statistical properties are unknown. Our nonlinear least squares approach does not require specific tuning and has also proved to be robust and reliable  for experimentally acquired data fitted with a realistic signal model.\\
Balanced GRE sequences are sensitive to off-resonance variations, requiring not only an extension of the model to include $\omega$ but also extra attention to the topology of the minimization landscape. We have analyzed the effect of time-varying RF excitations on the transient behavior of the signal and we have made a connection between the smoothness of the obtained response and the regularity of the objective function. In particular, smooth response throughout the whole sequence gives rise to a nicely shaped, basin-like optimization landscape. Moreover, the inherently low-frequency temporal response allows for an easier separation between the true signal component and the large high-frequency  artifacts caused by the randomized $k$-space undersampling. This is analogous to the reconstruction of strongly undersampled data in the compressed sensing framework \cite{cs} where the separation between data and artifacts is made possible by the incoherence between the sampling and sparsifying operators.\\
The transient analysis of the low-pass sequences and techniques borrowed from experimental design theory \cite{pukelsheim} lead to an algorithm which simultaneously enforces smoothness of   response and high-sensitivity to the parameters of interest. As a result, the iteratively solved nonlinear least squares fitting is successful and returns accurate parameter maps also in the case of highly undersampled experimentally acquired data. \\
The low-pass sequence obtained upon solution of Eq. (\ref{design}) exhibits a relatively high RF power level, which is caused by the large tip angles required. In this work, the constraints on the tip angle amplitude was set to  $180\dg$ but it could be adapted to a lower value if needed. For example, scanning at higher-field strengths such as 3T and beyond, requires a more conservative power management. As a supplementary experiment, we designed a low-pass sequence by solving problem ($\ref{design}$) with $\theta_{\max}=90\dg$. The sensitivity to $T_1$ and $T_2$ was lower than the one obtained with large tip-angle sequence but the dictionary-free reconstructions from the experimentally acquired data are still satisfactory (see the Supplementary Material).\\
 With our current desktop PC Matlab implementation, the computation time for the proposed method is, on average, about one second per voxel. This could be  accelerated by  more efficient software and hardware implementation, faster nonlinear least squares algorithms and Bloch equation simulators implemented on graphical processing units \cite{liu2017}. The aim of this work is to show that our dictionary-free approach is feasible. Acceleration strategies will be targeted in the future.\\
In this work we have considered balanced GRE sequences. Due to their superior SNR and robustness to motion, these sequences are used in a variety of applications. The signal model we employed makes use of one isochromat per voxel but it takes into account the effect of slice profile variation during the sequence. Our tests show that this model is accurate. Other signal models could be used, for instance, based on the extended phase graph formalism. This should not be a conceptual limitation for applying our method and could consequently also be targeted to design and reconstruct unbalanced sequences \cite{FISP_MRF}. 
\section{Conclusion}     
We propose to reconstruct multi-parametric maps from MR fingerprinting data by solving a nonlinear least-squares problem with established gradient-based minimization algorithms. By analyzing the relationship between the transient behavior of balanced MRF sequences and the corresponding optimization landscape of the reconstruction problem, we derived an experimental design strategy which leads to a very-well behaved objective function. Simulated and experimental tests show that such a dictionary-free approach combined with a specific experimental design strategy lead to correct MRF reconstructions. 
\section{Acknowledgment}
Part of this work was funded by the Dutch Technology Foundation (NWO-STW), grant number 14125.  \\
The authors are grateful to Dr. Tristan van Leeuwen and Dr. Chris Stolk for fruitful discussions and to Mrs Ying Lai Green for proofreading the manuscript.     
\ifCLASSOPTIONcaptionsoff
  \newpage
\fi

\section{Supplementary Material}
\begin{figure*}[t]
\centering
\epsfig{file=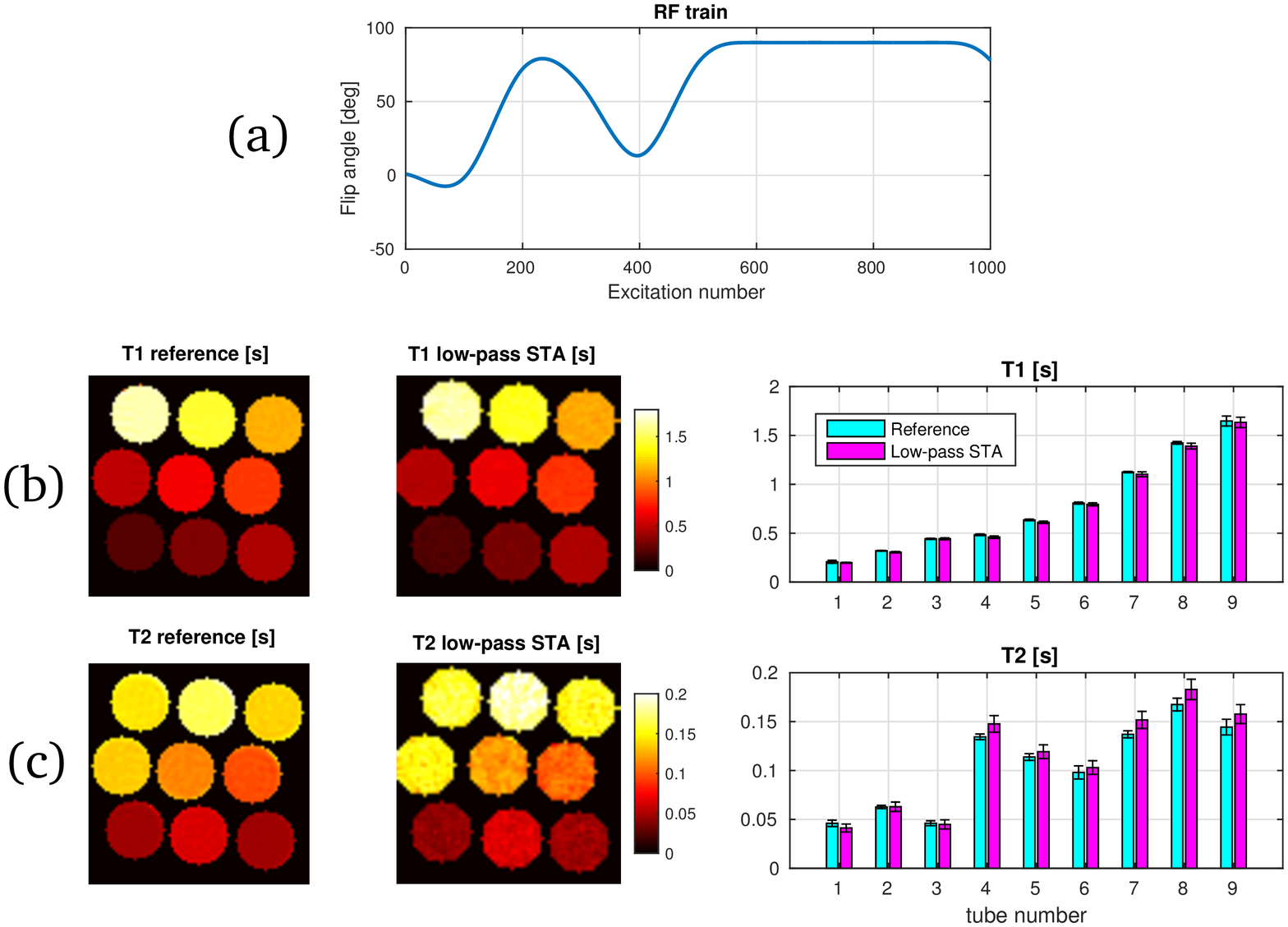, width=15cm}
\caption{\textbf{Supplementary Material:} Dictionary-free reconstructions for a small tip angle (STA) low-pass sequence designed according to Eq. (6) with $\theta_{\max}=90\dg$. (a): the RF train. (b-c) the $T_1$ and $T_2$ reconstructions from the 21 spokes radial acquisition of nine gel tubes with a 1.5 T MRI system. The reference data refers to an inversion-recovery spin-echo experiment. }
\label{STA}
\end{figure*}


\end{document}